\documentclass[twocolumn,floats,showpacs,amsmath,amssymb,prb]{revtex4}

\usepackage{graphicx}
                                                                                   
\begin{document}

\title{A path-integral molecular dynamics simulation of diamond}
\author{Rafael Ram\'{\i}rez}
\author{Carlos P. Herrero}
\affiliation{Instituto de Ciencia de Materiales,
         Consejo Superior de Investigaciones Cient\'{\i}ficas (CSIC),
         Campus de Cantoblanco, 28049 Madrid, Spain }
\author{Eduardo R. Hern\'andez}
\affiliation{Institut de Ci\`encia de Materials de Barcelona (ICMAB),
         Consejo Superior de Investigaciones Cient\'{\i}ficas (CSIC),
         Campus de Bellaterra, 08193 Barcelona, Spain }
\date{\today}

\begin{abstract}
Diamond is studied by
path integral molecular dynamics simulations of the
atomic nuclei in combination with 
a tight-binding Hamiltonian
to describe its electronic structure and total energy.
This approach allows us to quantify the influence of quantum zero-point 
vibrations and finite temperatures on both the
electronic and vibrational properties of diamond.
The electron-phonon coupling mediated by the zero-point
vibration reduces the direct electronic gap of diamond by 10$\%$.
The calculated decrease of the direct gap with temperature 
shows good agreement with the experimental data
available up to 700 K.
Anharmonic vibrational frequencies of the crystal have been obtained 
from a linear-response approach based on the path integral formalism.
In particular, the temperature dependence of the zone-center optical phonon 
has been derived from the simulations. 
The anharmonicity of the interatomic potential produces a red shift of
this phonon frequency.
At temperatures above 500 K, this shift is
overestimated in comparison to available experimental data.
The predicted temperature shift of the elastic constant $c_{44}$ displays
reasonable agreement with the available experimental results.

\end{abstract}


\maketitle

\section{Introduction}

Tetrahedral semiconductors such as silicon, germanium or diamond,
have served as model materials to study the electronic and vibrational
properties of crystals.
In particular, the effects of the lattice vibrations on the electronic 
properties, through the mechanism of electron-phonon coupling, have been
experimentally investigated by measuring the temperature dependence
of their optical excitation spectra. \cite{ca05a}
Furthermore, the observation of the 
dependence of such spectra with isotopic mass
has provided detailed information on the electronic
properties of these solids.
\cite{ca05b}  
Besides the electron-phonon interaction, the anharmonicity of lattice 
vibrations has been observed by the dependence of the phonon frequencies and    
line-widths with temperature and isotopic composition. \cite{ca05b} 

From a theoretical point of view, in spite of the impressive progress of 
ab initio methods for the investigation of the electronic structure of 
solids,
the atomic nuclei are usually considered either as fixed in their
crystallographic positions or by approximating their dynamics with 
classical mechanics. \cite{ca85}
Thus, the effects of the electron-phonon interactions on electronic 
properties and the effect of zero-point anharmonicity 
in the vibrational properties
of the lattice are usually neglected in these calculations.
As it has been pointed out in Ref. \onlinecite{ca05a},
these effects may be even larger that the error assumed
in the electronic ab initio calculations.

The electron-phonon interaction in tetrahedral semiconductors
has been studied theoretically by perturbation theory.
\cite{al81,zo92}
Within this approach, the reduction of the direct electronic gap 
due to zero-point vibrations of the lattice phonons is predicted
to be of 0.62 eV in diamond and of 0.06 eV in germanium.
Both energy shifts represent roughly a 10$\%$ fraction 
of the corresponding energy gaps, i.e.,
they are so large that a quantitative description of the electronic
structure can not be expected by a theory that neglects such effects.
Anharmonic shifts of the phonon modes of diamond and silicon
have been determined by combining density-functional 
perturbation theory with a frozen-phonon approach,\cite{la99}
and the results show good agreement with the experimental data available
from first-order Raman spectra. \cite{he91,sc97}
A review of the current status of lattice-dynamical calculations 
using density-functional perturbation theory  can be found in Ref.
\onlinecite{ba01}.

The path integral (PI) formulation of statistical mechanics
offers an alternative way to study finite temperature
properties that are related to the quantum nature of the
atomic nuclei. \cite{fe72,ce95}
Thus, the combination of the path integral formulation 
with electronic structure methods is an interesting alternative
to perturbational approaches for the study of 
electronic and vibrational properties of solids.
An advantage of this approach is that both the electrons and the atomic
nuclei are treated quantum mechanically in the framework of the
Born-Oppenheimer (BO) approximation, so that anharmonic and 
temperature effects can be evaluated
for both vibrational properties and the electronic structure. 
This unified scheme has been applied so far to the study of
solids and molecules containing light atoms.
\cite{marx96,tu97,ra98,si01,mo01,ch03,sa04,oh04}
 
In this paper we present a path integral molecular dynamics 
study of diamond at temperatures between 100 and 1200 K.
The electronic structure has been treated by a non-orthogonal
tight-binding (TB) Hamiltonian as a reasonable compromise to
reduce the computational cost of deriving the BO surface for the 
nuclear dynamics. 
In particular, we are interested in the investigation of 
electronic properties that are determined by the electron-phonon
coupling, as the dependence of the electronic gap with
temperature and isotopic mass.
Also vibrational properties that depend on the anharmonicity of
the interatomic potential will be studied with the help
of a linear response approach recently developed within
the path integral formalism.\cite{ra01} 
PI simulations of diamond using effective interaction potentials 
have been carried out earlier to study structural and thermodynamic
properties of this material.\cite{he01}

This paper is organized as follows. In Sec.\,II, we describe the
computational method and the models employed in our simulations.
Our results are presented and discussed in Sec.\,III, dealing with the 
direct electronic gap, the vibrational energy of the solid,
and the temperature dependence of the frequency of the optical phonon at the
center of the Brillouin zone (BZ) and the elastic constant $c_{44}$
of diamond.
In Sec.\,IV, we present the main conclusions of the paper.

\section{Computational Method}

\subsection{Simulation details}

The formalism employed here for the quantum treatment of 
electrons and nuclei is based on the combination of the
path integral formulation, to derive properties
of the atomic nuclei in thermal equilibrium,
with an electronic Hamiltonian to derive the
BO energy surface, $E_{BO}({\bf R})$, 
as a function of the nuclear configuration
${\bf R}$.
The path integral and electronic structure parts of
the resulting algorithm appear as independent blocks,
since the only electronic result required for the path integral
simulation is the value of the function $E_{BO}({\bf R})$, and possibly
its derivatives with respect to ionic positions. 
Thus, the combination of path integrals with
any chosen electronic Hamiltonian is straightforward.
For the present investigation of diamond we have chosen 
an efficient tight-binding one-electron effective Hamiltonian,
based on density functional (DF) calculations.\cite{po95}
The use of a simplified electronic Hamiltonian is a reasonable compromise
to explore the efficiency and capability of this unified formalism for the 
evaluation of electronic and vibrational properties 
of solids at finite temperatures.
The implementation of density functional or Hartree-Fock based
self-consistent methods is left for future development.  
The capability of tight-binding methods to simulate different properties of
solids and molecules has been reviewed by Goringe {\em et al.}\cite{go97}

The computational advantage of using the path-integral formulation of 
statistical mechanics is formulated
by the so-called "quantum-classical" isomorphism.
Thus, this method exploits the fact that
the partition function of a quantum system is formally equivalent to
that of a classical one, obtained by replacing each quantum particle 
(here, atomic nucleus) by a
ring polymer consisting of $L$ ``beads'', connected by harmonic
springs.\cite{gi88,ce95,fe72,kl90}
In many-body problems, the configuration space of the classical
isomorph is usually sampled by
Monte Carlo or molecular dynamics (MD) techniques. 
Here, we have employed the PI MD method, 
which has been found to require less computer time resources
in our problem.
Effective algorithms to perform PI MD simulations in the canonical $NVT$
ensemble have been described in detail by Martyna {\em et al.}\cite{ma96} 
and Tuckerman.\cite{tu02}
All calculations presented here were carried out in the canonical ensemble,
using originally developed MD software,
which enables efficient PI MD simulations on parallel supercomputers.

Simulations were carried out on a $2\times2\times2$ supercell of the
diamond face-centered cubic cell with periodic boundary conditions,
containing $N=$ 64 C atoms. 
The atomic mass of carbon was set to 12 amu.
The convergence of the internal energy has been checked
for some selected atomic configurations, by considering sets
of 1, 4 and 32 ${\bf k}$ points in the BZ 
of the simulation supercell.
The main effect of the ${\bf k}$ point sampling is found to be
a constant shift of the internal energy.
This systematic error is largely reduced 
in the calculation of properties obtained
as energy differences (e.g., energy shifts as a function
of temperature).
For this reason, we have chosen to use only the
$\Gamma$ point for the sampling of the BZ
of the simulation supercell.
A set of 4 ${\bf k}$ points increases the computer time
by a factor of 10 with respect to the $\Gamma$ point sampling,
without significant changes of the results presented here.

\begin{figure}
\includegraphics[width=12cm]{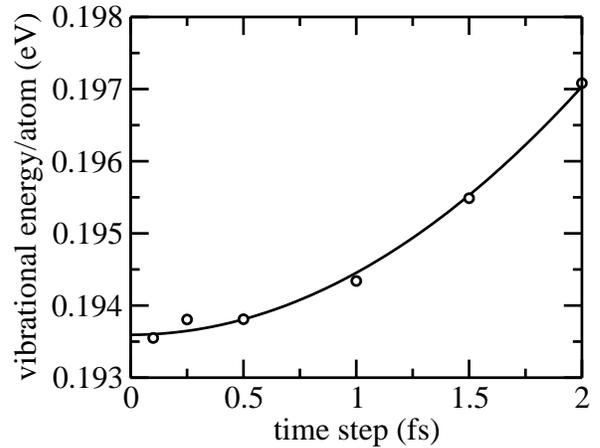}
\vspace{-9.6cm}
\caption{
Vibrational energy of diamond as a function of the time step, $\Delta t$,
employed in the PI MD algorithm. The results are derived at 300 K
with a Trotter number $L=20$.
The continuous line is a quadratic fit to the simulation results.
}
\label{f1}
\end{figure}

\begin{figure}
\includegraphics[width=11cm]{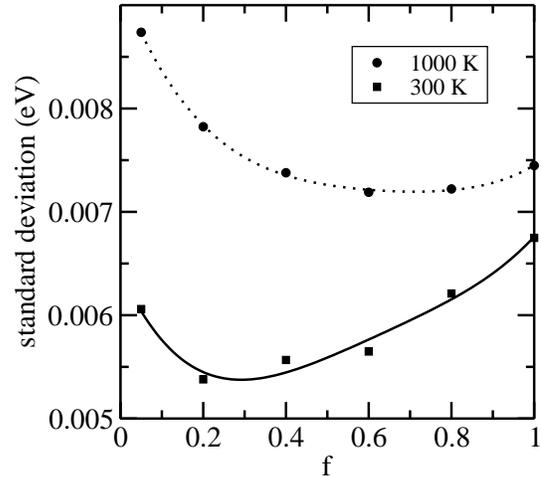}
\vspace{-8.1cm}
\caption{
Standard deviation of the total energy of the diamond supercell
as a function of the
scale factor $f$, used in Eq.~(\ref{Q}) to define the thermostat mass, $Q$.
The results correspond
to simulation lengths of $10^5$ MD steps at 300 K and $2\times10^5$
MD steps at 1000 K.
The lines are guides to the eye.
}
\label{f2}
\end{figure}

The simulation-cell parameter employed in our calculations is
taken from experimental data, and ranged from
7.1330 \AA\, at 100 K to 7.1552 \AA\, at 1200 K.\cite{sk57}
For a given temperature, a typical run consisted of $10^4$ MD steps for
system equilibration, followed by $5 \times 10^5$ steps 
for the calculation of
ensemble average properties.
To have a nearly constant precision in the path integral results
at different temperatures, we have taken a number of 
beads, $L$ (Trotter number), that scales with the inverse temperature
such that $LT = 6000$ K. 
For comparison with the results of our PI MD simulations, we have carried
out some classical MD simulations with the same interatomic interaction
(setting $L$ = 1).
The quantum simulations were performed using a staging transformation
for the bead coordinates.
Chains of four Nos\'e-Hoover thermostats 
were coupled to each degree of freedom to generate the canonical ensemble.
\cite{tu98}
To integrate the equations of motion we have used
the reversible reference system propagator algorithm (RESPA), which allows
one to define different time steps for the integration of the fast and slow
degrees of freedom.\cite{ma96}
For the evolution of the fast dynamical variables, that include the
thermostats and harmonic bead interactions, we used a
time step $\delta t = \Delta t/4$, where
$\Delta t$ is the time step associated to the calculation of DF-TB forces.
The convergence of the total energy as a function of $\Delta t$
is shown in Fig. \ref{f1}. 
A value of $\Delta t$ = 0.5~fs is found to provide 
adequate convergence.
We have also explored the convergence of the simulation as a
function of the thermostat mass, $Q$,
\begin{equation}
Q = \frac{f\beta \hbar^2}{L}  \,
\label{Q}
\end{equation}
where $\beta$ = $(k_B T)^{-1}$ is the inverse temperature,
and $f$ is a scaling factor.
The standard deviation of the total energy, as
derived from a block analysis,\cite{ca89} is displayed
in Fig. \ref{f2} at two different temperatures.
We observe that at 300 K the standard deviation can
be reduced by about 20 $\%$ by changing the $f$ factor from a value
of 1 to a value of 0.2.
Taking into account that the standard deviation varies with the number of 
simulation steps, $M_S$, as $M^{-1/2}_S$,
then, for a given threshold accuracy, a simulation run using
$f=0.2$ at 300 K requires 35 $\%$ less simulation steps 
than a run using $f=1$. 
In the simulations presented below, we have varied the 
parameter $f$ linearly for  
temperatures between 300 and 1000 K. The $f$ values changed
from $f=0.2$ (for $T \le 300$ K) to $f=1$ (for $T \ge 1000$ K).

\subsection{Calculation of anharmonic vibrational frequencies}
 
To calculate vibrational frequencies we will employ a method based
on the linear response (LR) of the system to vanishingly small forces applied
on the atomic nuclei.
With this purpose, we consider a LR function, the static isothermal
susceptibility $\chi^T$, that is readily derived from PI MD simulations of the
equilibrium solid, without having to explicitly impose any external forces 
during the simulation. This approach represents a significant improvement as
compared to the standard harmonic (HA) approximation.\cite{ra01}
A sketch of the method is given in the following.
 
Let us call ${\bf \{R}_p\} = \{x_{ip}\}$ the set 
of $3NL$ Cartesian coordinates of the
beads forming the ring polymers in the simulation cell
($i = 1, \dots, 3N; p = 1, \dots, L$).
We consider the set $\{X_i\}$ of centroid coordinates, with $X_i$
defined as the mean value of coordinate $i$ over the corresponding polymer:
\begin{equation}
X_i = \frac{1}{L}  \sum_{p=1}^L x_{ip}   \;.
\end{equation}
Then, the linear response of the quantum system to small external forces
on the atomic nuclei is given by the susceptibility
tensor ${\chi}^T$, which can be
defined in terms of centroid coordinates as\cite{ra01}
\begin{equation}
\chi^T_{ij} = \beta \sqrt{m_i m_j} \; \mu_{ij}  \;,
\label{chi_3d}
\end{equation}
where $\beta$ = $(k_B T)^{-1}$,
$m_i$ is the mass of the nucleus associated to coordinate $i$,
$\mu_{ij} = \langle X_i X_j \rangle -
             \langle X_i \rangle \langle X_j \rangle$
is the covariance of the centroid coordinates $X_i$ and $X_j$,
and $\langle\dots\rangle$ indicates an ensemble average along an MD run.
 
The tensor ${\chi}^T$ allows us to derive a LR approximation
to the low-lying excitation energies of the vibrational system,
that is applicable even to highly anharmonic situations.
The LR approximation for the vibrational frequencies reads
\begin{equation}
\omega_{n,LR} = \frac{1}{\sqrt{\Delta_n}}   \;,
\end{equation}
where $\Delta_n$ ($n = 1, \dots, 3N$) are eigenvalues of
$\chi^{T}$, and the LR approximation to the low-lying excitation energy of
vibrational mode $n$ is given by $\hbar\omega_{n,LR}$.
More details on the method and illustrations of its ability for predicting
vibrational frequencies of solids and molecules can be found
elsewhere\cite{ra01,ra02,lo03,ra05}.
In connection with the vibrational modes that actually appear in our
calculations, we note that the application of periodic boundary conditions
is physically equivalent to the consideration of lattice vibrations only at
the center (${\bf k}$ = {\bf 0}) of the BZ of the employed 
{\it simulation cell}.    
Modes with ${\bf k} \ne {\bf 0}$ violate the
periodic boundary conditions, because all
atomic images of an atom have distinct displacements, 
whose amplitude is modulated by both the propagation vector ${\bf k}$ 
of the vibrational mode and the translational vector of the image. 
However, periodic boundary conditions implies that all atomic
images must display a displacement identical to that one
of the atom located in the simulation cell,
a condition that is only met if the propagation vector is
${\bf k}$ = {\bf 0}.\cite{ra05}

\subsection{Calculation of one-electron energies} 

For the sake of clarity, we use the Schr\"odinger formulation
to derive the expectation value of electronic observables.
However, the final result is obtained in a form appropriate
to the path integral formulation.
Within the adiabatic Born-Oppenheimer approximation,\cite{fi89} 
the total wave function is written as
\begin{equation}  
\Psi_i({\bf r,R}) = \chi_i({\bf R}) \varphi_0({\bf r,R}) \;,
\label{phi}
\end{equation}
where $({\bf r,R})$ are the electronic and nuclear coordinates,
$\chi_i$ labels the nuclear wave function, and $\varphi_0$ represents
the electronic ground state configuration,
which depends parametrically on ${\bf R}$. 
Let us call $\epsilon_i$ the energy of the state $\Psi_i$.
Then, the canonical partition function, $Z$, is defined as
\begin{equation}
Z = \sum_i e^{-\beta\epsilon_i} \; .
\end{equation} 
We consider an electronic observable represented by the
operator  ${\cal E}({\bf r,R})$, that is a function of the
electronic coordinates and depends  parametrically on ${\bf R}$. 
Its canonical average,
$\langle E \rangle$, is defined as
\begin{equation}
\langle E \rangle = Z^{-1} \sum_i e^{-\beta\epsilon_i} 
                           \int d{\bf R} \int d{\bf r}
                           \Psi_i^*({\bf r,R}) {\cal E}({\bf r,R}) 
                           \Psi_i  ({\bf r,R})  \;.
\label{ave}
\end{equation} 
Now, we write this equation in an alternative way, better 
adapted to the path integral formulation.
First,
the function $E({\bf R})$ is defined as the expectation value of the operator
${\cal E}({\bf r,R})$ over the electronic wave function,
\begin{equation}
E({\bf R}) = \int d{\bf r} \varphi_0^*({\bf r,R})  {\cal E}({\bf r,R}) 
                           \varphi_0  ({\bf r,R})  \; .
\end{equation} 
The second definition is the funtion $\rho({\bf R},{\bf R})$,  
representing the diagonal elements of the normalized canonical
density matrix for the nuclear coordinates,
\begin{equation}
\rho({\bf R},{\bf R}) = Z^{-1}  \sum_i e^{-\beta\epsilon_i}  
                                | \chi_i({\bf R}) |^2 \; .
\end{equation}
Considering the factorization of $\Psi_i({\bf r,R})$ in Eq.~(\ref{phi}),
and using the last two definitions, we can rewrite the 
average $\langle E \rangle$ in Eq.~(\ref{ave}) as
\begin{equation}
\langle E \rangle = \int d{\bf R}\,\rho({\bf R},{\bf R}) E({\bf R}) \, .
\label{ave_pi}
\end{equation}
The last equation shows that electronic observables, $\langle E \rangle$, 
are obtained as ensemble averages over the nuclear configurations, ${\bf R}$, 
accessible in thermal equilibrium.
This equation can be readily used in combination with the
path integral sampling of the density matrix $\rho({\bf R},{\bf R})$.
We will apply it to calculate the canonical average,
$\langle E_n \rangle$, of
the $n$-th energy eigenvalue of the electronic Hamiltonian.
For convenience, let us define the following probability density
for the one-electron state $E_n$,
\begin{equation}
\rho_n(E) = \int d{\bf R}\,\rho({\bf R},{\bf R})\,\delta(E_n({\bf R})-E) \, ,
\label{rho}
\end{equation}    
where $\delta$ is the Dirac delta function. 
$\rho_n(E)$ can be easily accumulated
during a PI~MD simulation.
The expectation value, $\langle E_n \rangle$, as given by  
Eq.~(\ref{ave_pi}), can now be written as the
first moment of the distribution $\rho_n$
\begin{equation}
\langle E_n \rangle = \int dE \rho_n(E) E \, .
\label{moment}
\end{equation}
The direct electronic gap of diamond is derived as 
\begin{equation}
E_g = \langle E_c \rangle - \langle E_v \rangle  \, ,
\end{equation}  
where $E_c$ and $E_v$ are the 
one-electron states associated with the bottom of the
conduction band and the top of the valence band at the reciprocal point
${\bf k = 0}$.

\section{Results and Discussion}

In this section we present the main results derived from our PI MD
simulations of diamond as a function of temperature. Whenever possible
we will compare the simulation results to available experimental data. 
Results concerning the electronic and vibrational properties
are presented in the next subsections.

\subsection{Electronic properties}

\subsubsection{One-electron states}

\begin{figure}
\includegraphics[width=9.6cm]{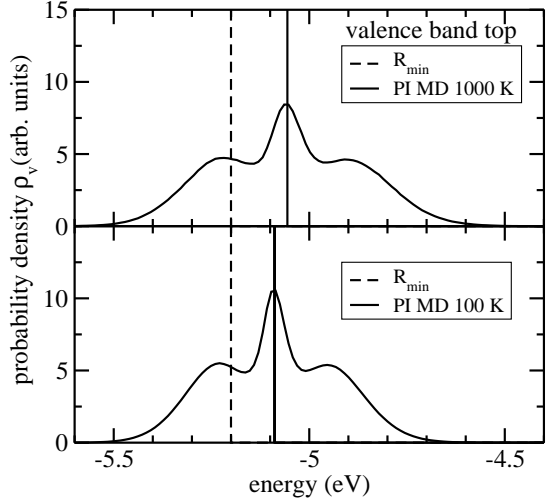}
\vspace{-5.6cm}
\caption{
Probability density functions for the valence band top of diamond
obtained by PI MD simulations at 100 and 1000 K.
The expectation value of the electronic level,
$\langle E_v \rangle$, is shown as a vertical
continuous line.
The broken line shows the valence band top,
$E_v$, for the nuclear configuration
${\bf R_{min}}$, with the atoms fixed at their crystallographic positions
and the cell parameter fixed at the equilibrium value at 100 K.
}
\label{f3}
\end{figure}

The top of the valence band, $E_v$, for a diamond crystal with the atoms fixed 
in their crystallographic positions, ${\bf R_{min}}$,  
is three-fold degenerate.
Each of these three one-electron levels leads 
to identical probability densities,
$\rho_v$, as defined in Eq.~(\ref{rho}). 
The function $\rho_v(E)$, derived from our PI MD simulations, is represented  
in Fig. \ref{f3} at two temperatures, 100 and 1000 K.  
The probability density shows three distinct maxima 
reflecting that the underlying electronic state
is three-fold degenerate. 
In each case, the continuous vertical line displays 
the expectation value of the
valence band top, $\langle E_v \rangle$, defined as the 
first moment of the distribution $\rho_v$ [see Eq.~(\ref{moment})].  
The temperature effect in $\langle E_v \rangle$ is a shift 
of 0.03 eV toward higher energies, when the temperature increases
from 100 K to 1000 K. 
This shift is a consequence of the different magnitude of
the electron-phonon interaction at both temperatures
(longer atomic displacements at higher $T$).
The dotted line in the figure represents the energy of the top 
of the valence band, $E_v$, for a crystal with atoms fixed at their equilibrium
positions and 
cell parameter set to the equilibrium value at $T =100$ K ($a=3.5665$ \AA).  
The PI MD simulation predicts that at 100 K the top of the valence 
band is shifted by 0.11 eV with respect to the result obtained for the
${\bf R_{min}}$ configuration.

\begin{figure}
\includegraphics[width=10cm]{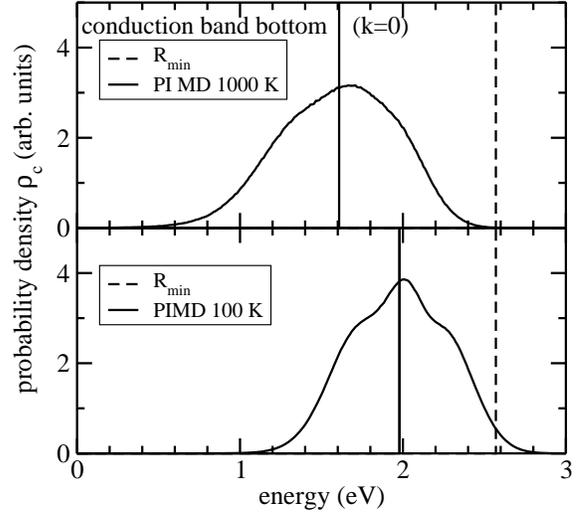}
\vspace{-5.6cm}
\caption{
Probability density function for the conduction band bottom
at the reciprocal point ${\bf{k = 0}}$, obtained by
PI MD simulations at 100 and 1000 K.
The expectation value for this electronic level, $\langle E_c \rangle$,
is shown by a continuous vertical line.
The broken line shows the position of $E_c$ for the static
nuclear configuration ${\bf R_{min}}$.
}
\label{f4}
\end{figure}

In Fig. \ref{f4} we present the results corresponding to 
the one-electron state $E_c$, i.e., the conduction band 
bottom at the reciprocal point ${\bf k=0}$.
The energy shifts found for $E_c$ as a consequence of the
electron-phonon interaction are larger and of opposite sign as those
encountered  for the valence band.
At 100 K we observe a downwards shift of $\langle E_c \rangle$ by about 
-0.59~eV with
respect to the ${\bf R_{min}}$ configuration. 
The temperature shift in $\langle E_c\rangle$ amounts 
to -0.37 eV~between 100 and 1000~K.

\subsubsection{Direct electronic gap}

\begin{figure}
\includegraphics[width=10cm]{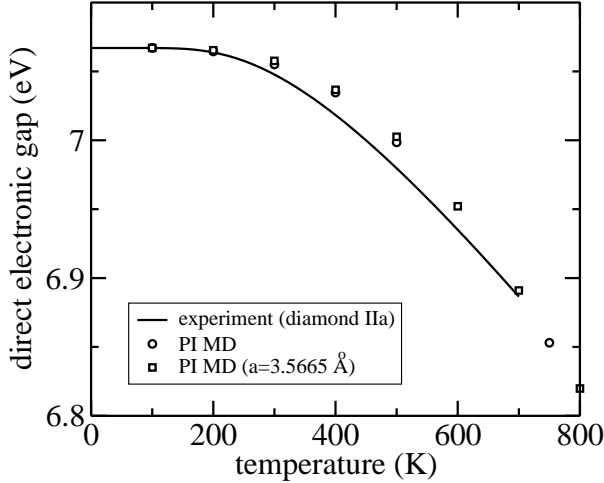}
\vspace{-6.0cm}
\caption{
Temperature dependence of the direct electronic gap of diamond
obtained by our PI MD simulations.
Open circles were derived
from simulations that take into account the thermal expansion of the
lattice.
Open squares correspond to simulations where the cell parameter was
fixed to the equilibrium value at 100 K.
The continuous line is the fit to the experimental data given in
Ref. ~\onlinecite{lo92} for diamond IIa.
}
\label{f5}
\end{figure}

The first direct gap, $E_g = \langle E_c \rangle- \langle E_v \rangle$, 
of diamond has been portrayed in Fig. \ref{f5} as a function
of temperature.
The open circles have been derived from PI MD simulations where the 
thermal expansion of the crystal lattice 
has been taken into account by varying the value of 
the cell parameter $a$. 
Open squares are PI MD results derived at different temperatures
with a cell parameter fixed at the equilibrium value 
at 100 K. 
We note that the effect of the thermal expansion on the
electronic gap is a slight reduction of the gap. 
The temperature dependence of the electronic gap predicted
by our PI MD simulations is in reasonable agreement with the
experimental results reported in Ref. \onlinecite{lo92} for diamond IIa
up to 700 K,
based on measurements of the complex dielectric
function by spectroscopic ellipsometry.
The slope of the simulation results at temperatures above 500 K
is larger than that of the experimental data.
A possible reason for this behavior is an
overestimation of anharmonic effects by the DF-TB potential model. 
Although we expected a reasonable agreement between experimental and
theoretical results at low $T$, the coincidence shown in Fig. \ref{f5}
for the absolute value of the first direct gap is fortuitous.
In fact, the displayed experimental values, derived from first-derivative
line-shape analysis of the complex dielectric function, are shifted
by about 0.07 eV toward higher energies, in case that
they are obtained by second-derivative
line-shape analysis.\cite{lo92}

\begin{figure}
\includegraphics[width=11cm]{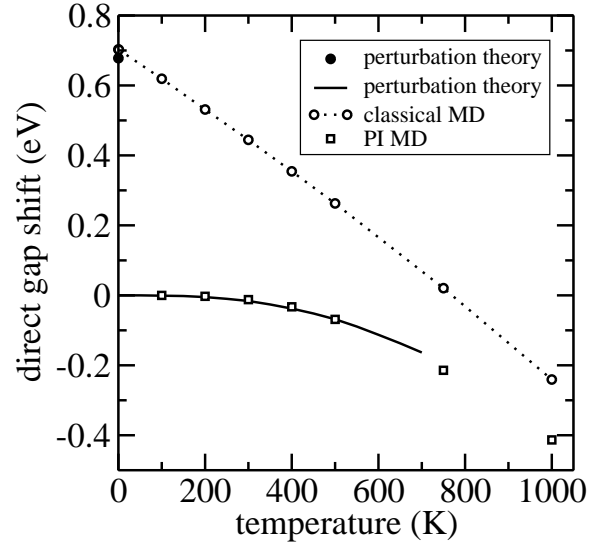}
\vspace{-6.7cm}
\caption{
Relative shifts of the direct electronic gap of diamond.
Open circles are results from classical MD simulations, while
open squares correspond to quantum PI MD simulations.
The solid circle at $T=0$ and the continuous line are results
presented in Ref. ~\onlinecite{zo92} based on perturbation theory.
For both PI MD simulations and perturbation theory, shifts are given
with respect to the corresponding quantum limit at $T=0$.
The dotted line is a guide to the eye.
}
\label{f6}
\end{figure}

To quantify the influence of nuclear quantum effects on the 
value of the direct electronic gap of diamond we have performed a series of
classical MD simulations as a function of temperature.
The shifts of the energy gap obtained in the classical simulations
are compared to the PI MD results in Fig. \ref{f6}.
The most prominent quantum effect appears in the low temperature limit
as a consequence of the zero-point vibration.
The renormalization of $E_g$ amounts to 0.7 eV at $T=0$. 
This value agrees well with the perturbational
treatment of the electron-phonon coupling in Ref. \onlinecite{zo92}, 
whose result is represented by a closed
circle in Fig. \ref{f6}.
We stress that the zero-point renormalization of $E_g$ amounts to
about 10$\%$ of its value. 
The PI MD results show satisfactory agreement with
the perturbation theory data available up to 700 K.\cite{zo92}
The main discrepancy found between both sets of results
is that the slope at temperatures above 500 K is
larger for PI MD than for perturbation theory.
The overestimation of anharmonic effects by the  DF-TB model
is a probable explanation for this behavior.
Differences between the quantum and classical results for $E_g$ are
significant in the whole studied temperature range.
In particular, at room temperature the classical result deviates from
the PI MD data by about 0.45 eV.

We have calculated the shift of the direct electronic  gap of diamond
as a function of the isotopic mass at 300 K.
The calculated energy gap for $^{12}$C amounts to 7.054 eV.
For $^{13}$C, this gap increases 
to 7.081 eV at the same temperature.
The isotopic effect of 27 meV at 300 K is in reasonable
agreement to the value reported in Ref. \onlinecite{zo92}
of 22 meV, based on perturbation theory in the zero temperature limit.

\subsection{Vibrational Properties}

\subsubsection{Vibrational energy}

\begin{figure}
\includegraphics[width=11cm]{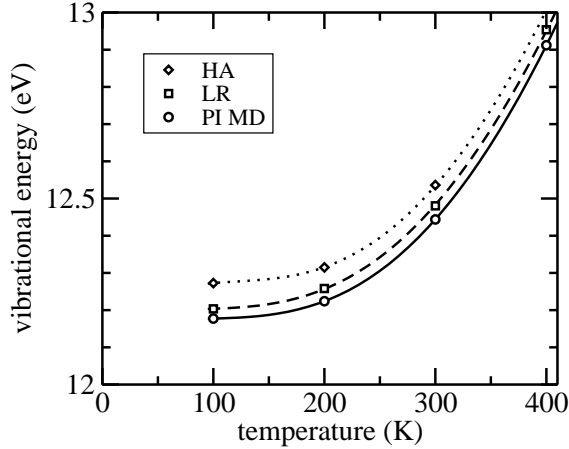}
\vspace{-8cm}
\caption{
The vibrational energy of the simulation cell of
diamond as a function of temperature is shown by open circles.
The zero of energy corresponds to the crystal with fixed atoms
and cell parameter $a=3.5665$ \AA.
The harmonic vibrational energy  derived from the set
of harmonic and linear response frequencies is displayed as
open diamonds and squares, respectively.
The lines are guides to the eye.
}
\label{f7}
\end{figure}

The vibrational energy of the simulation cell 
of diamond, as derived by the PI MD simulations,
is presented as a function of temperature in Fig. \ref{f7} (solid circles).
The zero of the energy scale corresponds to a diamond crystal
with fixed atoms and with the cell parameter fixed 
at the equilibrium value at 100 K. 
The thermal occupation of excited vibrational states is 
evident in Fig. \ref{f7} by the increase of the vibrational
energy with temperature.
To quantify the anharmonic effect on the vibrational energy of the
crystal, we have plotted in Fig. \ref{f7} the harmonic vibrational energy
(solid diamonds).
The set of harmonic frequencies has been 
derived by diagonalizing the dynamic matrix in a simulation cell
with the experimental equilibrium lattice parameter at each temperature.
At the lowest studied temperature the harmonic result deviates
from the PI MD value by about 0.1 eV, which amounts to about
0.8 \% of the total vibrational energy.
This error of the harmonic approximation is a consequence of the
anharmonicity of the phonon vibrations.

The set of anharmonic vibrational frequencies, $\omega_{n,LR}$, derived by our
linear response approach, are expected to represent an improved
description of the vibrational problem.
Thus, we have recalculated the harmonic vibrational energy
by considering $\omega_{n,LR}$ as a set of 
renormalized phonon frequencies.
The result is shown as open squares in Fig. \ref{f7}.
Most of the error of the harmonic approximation is corrected by
the LR frequencies.
At low temperatures the absolute error of 
the improved estimation of the vibrational energy 
amounts to 0.03 eV, i.e., about 0.2 \% of the PI MD result.
Our conclusion from this comparison is that the set of
LR frequencies provides a consistent description of 
anharmonic effects in the employed DF-TB model, 
in line with previous
results of vibrational properties on molecular and solid
state systems.\cite{ra01,lo03,ra05}
In the following, we focus on the study of two particular LR phonons
that can be compared to available experimental data.

\subsubsection{Optical phonon at ${\bf k=0}$}

The highest energy phonon of the diamond crystal is the 
optical phonon at the center of the BZ (${\bf k=0}$).  
At 200 K, the LR wavenumber of this phonon amounts to
1396 $\pm$ 5 cm$^{-1}$.
The harmonic result, obtained for the equilibrium cell
parameter at this temperature, is 1407 cm$^{-1}$. 
The difference between the LR and harmonic result is a consequence
of the anharmonicity of the interatomic potential that
induces a slight softening in the phonon frequency.
The optical phonon wavenumber of diamond in the zero temperature limit
is found in the first-order Raman spectrum at 1335 cm$^{-1}$.\cite{he91}
The DF-TB potential overestimates the experimental wavenumber 
by about 61 cm$^{-1}$.
This error is lower than that found in other
tight-binding parametrizations.\cite{wa90,ko97}

\begin{figure}
\includegraphics[width=10cm]{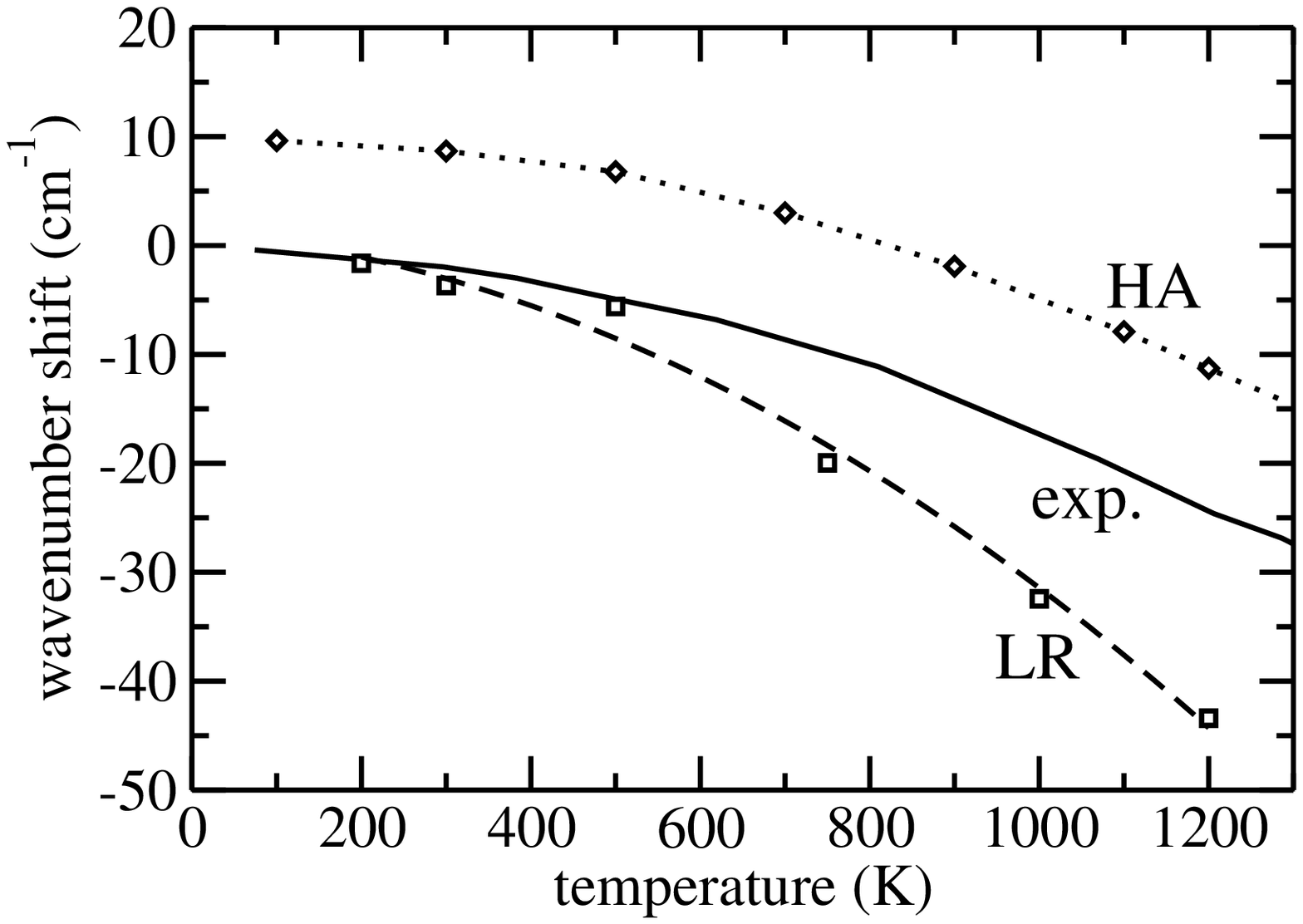}
\vspace{-7.0cm}
\caption{
Temperature shift of the optical phonon at $\Gamma$
(${\bf k=0}$) of diamond.
The results of the LR and harmonic approximations are compared with
a fit to the experimental data of Ref. \onlinecite{he91}.
The extrapolated LR value at $T=0$ has been
employed as zero of the vertical axis for both the LR and HA results.
The zero of the experimental data corresponds to the extrapolated
experimental wavenumber at $T=0$.
}
\label{f8}
\end{figure}

The relative shift of the optical phonon in diamond
is presented as a function of temperature in Fig. \ref{f8}.
The comparison between the harmonic and LR results shows
that anharmonic effects lead to a softening of the phonon
mode in the studied temperature range up to 1200 K.
The LR results 
deviate from the experimental data at temperatures
above 500 K.
This deviation shows that the
anharmonicity of the optical phonon is overestimated by the DF-TB
Hamiltonian at those temperatures.
We have previously commented on the enhanced anharmonicity
of the DF-TB model in relation to the decrease of the direct
electronic gap at temperatures above 500 K.

\subsubsection{Elastic constant c$_{44}$}

The lowest energy phonon, $\omega_1$, determined by either the LR or HA
approximations is 12-fold degenerated in the employed simulation cell.  
The wave vector,
${\bf k_1}$, of this phonon state has been identified
as the midpoint between $\Gamma$ and $X$ points along
the $\Delta$[100] direction in reciprocal space,
with coordinates $(\pi/a,0,0)$.
This means that the phonon $\omega_1$ corresponds to the transverse 
acoustic (TA) branch.
The identification of the wave vector of $\omega_1$ has been
possible by solving the dynamical matrix at some selected ${\bf k}$ points
of the BZ of the primitive unit cell of diamond.
The phonon velocity, $v_{TA}$, along the TA branch, 
$\omega_{TA}(k_{TA})$, is defined as the slope
of the dispersion branch at the origin,
\begin{equation}
v_{TA} = \lim_{k_{TA}\to 0} \frac{\omega_{TA}}{k_{TA}} 
       \approx  c \frac{\omega_{1}}{k_{1}} \; .
\label{v}
\end{equation}
In this equation, $k_{1}=\pi/a$ is the modulus of the 
wave vector ${\bf k_1}$ 
associated to the phonon $\omega_{1}$.
The constant $c=1.037$ corrects, for the case of the HA approximation,
the finite difference error encountered by using the finite vector ${\bf k_1}$ 
to calculate the slope at the origin.
The elastic constant $c_{44}$ can be derived from $v_{TA}$ by the
relation
\begin{equation}
c_{44} = \rho v_{TA}^2 \; ,
\label{c}
\end{equation}
where $\rho$ is the density of the diamond crystal.

\begin{figure}
\includegraphics[width=10cm]{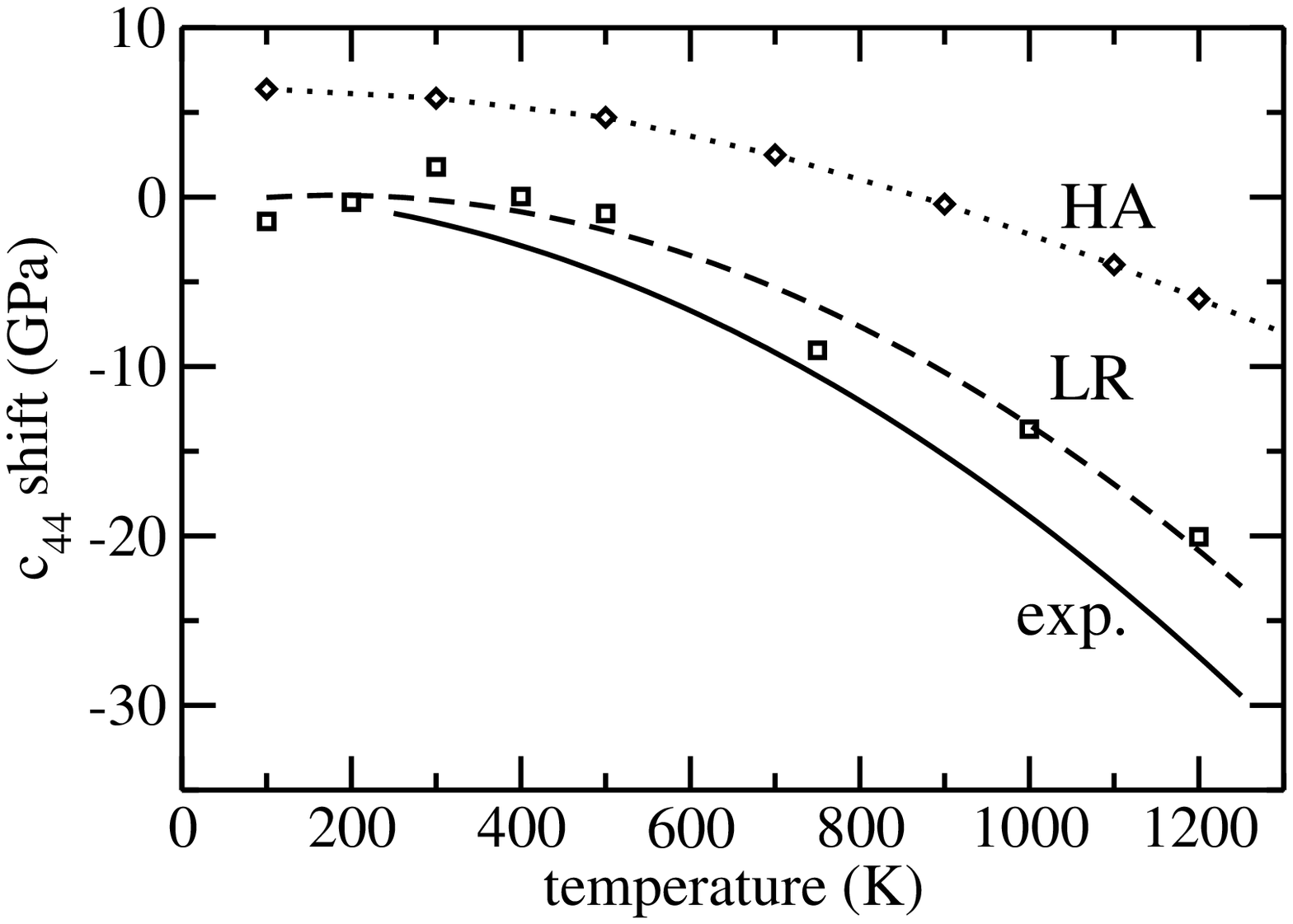}
\vspace{-7.5cm}
\caption{
Temperature shift of the elastic constant $c_{44}$ of diamond.
The results of the LR and harmonic approximations are compared to
a fit to the experimental data of Ref. \onlinecite{zo98}.
The extrapolated LR value at $T=0$ has been
employed as zero of the vertical axis for both the LR and HA results.
The zero of the experimental data corresponds to the extrapolated
experimental constant at $T=0$.
}
\label{f9}
\end{figure}

Using the harmonic value of $\omega_{1,HA}$ at 200 K
we obtain with Eqs.~(\ref{v}) and (\ref{c}) 
a value for $c_{44}$ of 551 GPa. 
The estimation of the elastic constant using the 
LR wavenumber, $\omega_{1,LR}$, in Eq.~(\ref{v})
is of 545  GPa.  
The experimental result derived from Brillouin scattering amounts to
576 GPa.\cite{zo98} 
The shift of the elastic constant $c_{44}$ 
with temperature is plotted in Fig. \ref{f9}.
The comparison of the LR and harmonic results shows that anharmonic
effects cause a reduction of the value of the elastic constant $c_{44}$ in the
whole studied temperature range.
The comparison to the experimental data shows that the
DF-TB model gives a reasonable prediction of the temperature
shift of the $c_{44}$ elastic constant, even though
there appears a systematic underestimation of this shift.

\section{Conclusions}

The simulations presented here for diamond are based on the    
treatment of electrons and nuclei as quantum particles in 
the framework of the Born-Oppenheimer approximation.
The use of the path integral formulation for the atomic
nuclei allows us to obtain the vibrational and electronic 
properties of the solid at finite temperatures.
We have chosen a simplified electronic Hamiltonian
to develop the algorithms required for the simulation of 
solid state systems, but this limitation should be
eliminated in the future by the implementation of 
improved electronic structure methods.

The temperature and  isotopic dependence
of the first direct gap of diamond predicted by our PI MD
simulation shows good agreement with the available experimental
results, based on spectroscopic ellipsometry,\cite{lo92}
and theoretical results, based on perturbation theory.\cite{zo92}
Thus, the employed simulation model has demonstrated its capability
to realistically describe electronic properties that are determined
by electron-phonon interactions.
The effect of the zero point vibrations of the lattice
phonons of diamond in its first direct gap is a reduction
of the gap by about 0.7 eV.        
This effect is so large that
any theoretical approach aiming
at a quantitative determination of the electronic gap of diamond
can not be based only on an improved solution of the many-body electronic
problem, but it should also 
include the treatment of the electron-phonon coupling. 

Anharmonic effects in the lattice vibrations have been
derived by a linear response approach based on the path integral
formulation.
This approach allows us to derive 
anharmonic vibrational frequencies from the
study of spatial correlations in the displacements 
of the vibrating nuclei in the simulation cell.
Anharmonic effects are responsible for a reduction of the
vibrational energy of the solid of about 1 \%, with respect to the
result predicted by a harmonic approximation.
The temperature shift of the optical phonon at ${\bf k=0}$
is larger that the experimental result determined by Raman
spectroscopy.\cite{he91}
We consider that this limitation is a consequence of the
parametrization of the employed DF-TB one-electron
Hamiltonian,\cite{po95} that overestimates the anharmonicity
of the highest frequency phonon of the lattice.
Better agreement is found in the comparison of the elastic constant
$c_{44}$ derived from the simulations with the
experimental values obtained by Brillouin scattering.\cite{zo98}
 
We plan to extend our simulations to more complex systems
like hydrogen impurities in diamond,\cite{he06} where the
presence of light impurities should strengthen further
the influence of quantum nuclear effects in the
electronic and vibrational properties of the lattice.

\begin{acknowledgments}
The calculations presented here were performed at the Barcelona
Supercomputing Center (BSC-CNS).  
This work was supported by CICYT through Grant
No. BFM2003-03372-C03-03 and by CAM through project S-0505/ESP/000237.
ERH thanks DURSI (regional government of Catalonia)
for funding through project 2005SGR683.
\end{acknowledgments}

\bibliographystyle{apsrev}

\end{document}